\documentclass[a4paper, 12pt]{article}
\usepackage{authblk}
\usepackage{sectsty}
\usepackage{cite}

\makeatletter
\renewcommand{\@biblabel}[1]{#1.}
\makeatother

\sectionfont{\large}

\begin{document}
\title{Entropy formula of N-body system}

\author{Jae Wan Shim}
\affil{Materials and Life Science Research Division,\\Korea Institute of Science and Technology, and\\Major of Nanomaterials Science and Engineering, KIST Campus,\\Korea University of Science and Technology, \\5 Hwarang-ro 14-gil, Seongbuk, Seoul 02792, Republic of Korea\\Correspondence: jae-wan.shim@kist.re.kr}

\date{}
\maketitle
\abstract
We prove a proposition that the entropy of the system composed of finite $N$ molecules of ideal gas is the $q$-entropy or Havrda-Charv\'at-Tsallis entropy, which is also known as Tsallis entropy, with the entropic index $q=\frac{D(N-1)-4}{D(N-1)-2}$ in $D$-dimensional space. The indispensable infinity assumption used by Boltzmann and others in their derivation of entropy formulae is not involved in our derivation, therefore our derived formula is exact. The analogy of the $N$-body system brings us to obtain the entropic index of a combined system $q_C$ formed from subsystems having different entropic indexes $q_A$ and $q_B$ as $\frac{1}{1-q_C}=\frac{1}{1-q_A}+\frac{1}{1-q_B}+\frac{D+2}{2}$. It is possible to use the number $N$ for the physical measure of deviation from Boltzmann entropy.
\\

\section{Overview}
Havrda-Charv\'at-Tsallis entropy or the $q$-entropy is a generalisation of Boltzmann entropy\cite{HC, Tsallis,Tsallis2009}. The first appearance of the formula of the $q$-entropy is on the paper of Havrda and Charv\'at for the purpose of information entropy and Tsallis has rediscovered it for Boltzmann-Gibbs statistical mechanics. The generalisation is characterised by a single parameter known as the entropic index $q$ and Boltzmann entropy is retrieved when $q$ approaches one. The endeavors to discover additional factors that explain the roles of the entropic index $q$ and to find empirical or approximated relations for the entropic index have been in various fields including physics, chemistry, and economics\cite{Cho, plastino, plastino1994, tsallis1996generalized, tsallis1999re, beck, walton, ion, upadhyaya, weinstein, borland, lutz, reynolds, reis, ausloos, vilar, liu}. However, Boltzmann derived his entropy formula by considering a very fundamental physical system composed of infinite number of ideal gases where the infinity was an indispensable constraint for using Stirling's approximation with respect to the factorial function that appeared in his procedure\cite{boltzmann}. Meanwhile, the derivation of the precise entropy formula for the Boltzmann's system of $N$ molecules seems unknown and hard to be accomplished by following his way because the infinity assumption is not allowed to use any more in this case. Here we show the entropy of the system composed of finite $N$ molecules of ideal gas is Havrda-Charv\'at-Tsallis entropy or the $q$-entropy with the entropic index $q=\frac{D(N-1)-4}{D(N-1)-2}$ in $D$-dimensional space. This analytically obtained result partially reveals the detail behind Boltzmann entropy concealed by using the infinity assumption. In addition, Havrda-Charv\'at-Tsallis entropy or the $q$-entropy is now established on a physical system which is as precise and fundamental as the foundation of Boltzmann entropy. By using the analogy of the $N$-body system, it is possible to obtain the entropic index of a combined system $q_C$ formed from subsystems having different entropic indexes $q_A$ and $q_B$ as $\frac{1}{1-q_C}=\frac{1}{1-q_A}+\frac{1}{1-q_B}+\frac{D+2}{2}$. The number $N$ can be used for the physical measure of deviation from Boltzmann entropy.

\section{Entropy formula of N-body system}
Let us consider $D$-dimensional space where molecules reside so that the velocities of $N$ molecules can be described in $D\times N$-dimensional vector space. The probability density function, established on Borel $\sigma$-algebra of the vector space equipped with Dirac measure, of an isolated system composed of $N$ molecules of ideal gas with respect to their velocities can be written by
\begin{equation}
f(\mathbf{v}_1,\cdots, \mathbf{v}_N)=c \; {\delta}\left(\sum_i \mathbf{v}_i^2-U\right)
\end{equation}
where $\mathbf{v}_i$ is the velocity vector of the $i$th molecule, $\mathbf{v}_i^2$ is the inner product between $\mathbf{v}_i$ and itself, $c$ is a normalising constant,  $\delta$ is the Dirac delta function, and $mU/2$ is a given kinetic energy of the system with $m$ being the mass of a molecule. This formula states that the velocity distribution is constrained by the kinetic energy of the system. By integrating $f(\mathbf{v}_1,\cdots, \mathbf{v}_N)$ with respect to $D(N-1)$ variables of the velocity components except those of $\mathbf{v}_j$ for any specific $j$, we can obtain the marginal probability $f(\mathbf{v}_j)$ or $f(\mathbf{v})$ with eliminating the subscript in $\mathbf{v}_j$ as
\begin{equation}
f(\mathbf{v})= \frac{\Gamma\left(\frac{DN}{2}\right)}{\Gamma\left(\frac{D(N-1)}{2}\right)(\pi U)^{\frac{D}{2}}}\left(1-\frac {\mathbf{v}^2} {U} \right)_+^{\frac {D(N-1)-2} {2}}
\end{equation}
for $N>1+\frac{2}{D}$ where $\Gamma$ is the gamma function also known as the Euler integral of the second kind and the positive symbol as a subscript of the parenthesis signifies that we take zero for the value of $f(\mathbf{v})$ when it is negative. The uses of the polar coordinate system and the surface area of the unit sphere in multi-dimensional vector space is helpful for obtaining the marginal probability\cite{Cercignani1969}.
  
On the other hand, following Havrda-Charv\'at-Tsallis entropy or the $q$-entropy formula\cite{Tsallis}, we write an entropy $S$ for a probability density function $p(\mathbf{x})$ where $\mathbf{x}$ is a random variable in $D$-dimensional space by
\begin{equation}
S= k \frac {1-\int p(\mathbf{x})^q d\mathbf{x}}{q-1} 
\end{equation}
where $k$ is the Boltzmann constant. Emphasizing that the normalised $q$-expectation\cite{Abe2006} is different from the ordinary expectation evaluated by $\int g(\mathbf{x}) p(\mathbf{x})d\mathbf{x}$ for a given function $g(\mathbf{x})$, we define the normalised $q$-expectation $E$ by
\begin{equation}
E[g(\mathbf{x})]=\frac {\int g(\mathbf{x}) p(\mathbf{x})^q d\mathbf{x}}{\int p(\mathbf{x})^q d\mathbf{x}}
\end{equation}
in which $E$ approaches the ordinary expectation when $q \rightarrow 1$. By maximising the entropy formula $S$ under the constraints of
$\int p(\mathbf{v}) d\mathbf{v} =1$ and $E[\mathbf{v}^2]=\frac{D(1-q)}{2+D(1-q)}U$, we obtain
\begin{equation}
p(\mathbf{v})=\frac {\Gamma\left(\frac{2-q}{1-q}+ \frac{D}{2} \right)}{\Gamma \left(\frac{2-q}{1-q}\right) (\pi U)^{\frac{D}{2}} } \left(1-\frac{\mathbf{v}^2}{U} \right)_+^{\frac{1}{1-q}}
\end{equation}
for $q<1$. Now, it is clear that by the following definition,
\begin{equation}
q=\frac{D(N-1)-4}{D(N-1)-2},
\end{equation}
we have demonstrated that the probability density function $p(\mathbf{v})$ obtained from the maximisation of the entropy $S$ under the constraints is identical to $f(\mathbf{v})$ evaluated from the description of the $N$-body system by the Dirac delta function. Consequently, we can establish the entropy formula of the $N$-body ideal gas system by using the entropy $S$.

\section{Entropic index of a combined system}
Let us consider two systems $A$ and $B$ having the entropic indices $q_A$ and $q_B$; and the number of ideal gas molecules $N_A$ and $N_B$, respectively. Then, by using the $N$-body analogy, the entropic index $q_C$ of the combined system $C$ of the systems $A$ and $B$ can be obtained by
\begin{equation}
\frac{1}{1-q_C}=\frac{1}{1-q_A}+\frac{1}{1-q_B}+\frac{D+2}{2}
\end{equation}
or, by using the number of ideal gas molecules as a physical measure,
\begin{equation}
N_C=N_A+N_B
\end{equation}
where $N_C$ is the number of ideal gas molecules of the combined system $C$.

\section*{Acknowledgment}
This work was partially supported by the KIST Institutional Program.
\bibliographystyle{naturemag}
\bibliography{ScientificReportsVer3}
\end{document}